\documentclass{article}

\usepackage{arxiv}
\usepackage{subcaption}
\usepackage[utf8]{inputenc}
\usepackage[T1]{fontenc}
\usepackage{hyperref}
\usepackage{url}
\usepackage{booktabs}
\usepackage{amsfonts}
\usepackage{nicefrac}
\usepackage{microtype}
\usepackage{lipsum}
\usepackage{graphicx}
\usepackage{natbib}
\usepackage{doi}
\usepackage[percent]{overpic}
\usepackage{xcolor}
\usepackage{multirow}
\usepackage{enumitem} 
\usepackage{appendix}
\usepackage{float}

\title{Learning from nature: insights into GraphDOP's representations of the Earth System}

\renewcommand{\shorttitle}{Learning from nature}

\hypersetup{
pdftitle={XXX},
pdfsubject={},
pdfauthor={},
pdfkeywords={},
}

\newbox{\orcid}\sbox{\orcid}{\includegraphics[scale=0.06]{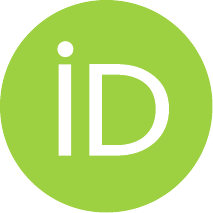}} 

\author{
    {\href{https://orcid.org/0000-0002-3662-5382}{\usebox{\orcid} Peter Lean}}
    \And {\href{https://orcid.org/0009-0007-7798-6524}{\usebox{\orcid} Mihai Alexe}}
    \And {\href{https://orcid.org/0000-0002-6070-2544}{\usebox{\orcid} Eulalie Boucher}}
    \And {\href{https://orcid.org/0000-0003-1869-3426}{\usebox{\orcid} Ewan Pinnington}}
    \And {\href{https://orcid.org/0000-0003-3952-586X}{\usebox{\orcid} Simon Lang}}
    \And {\href{https://orcid.org/0000-0003-2808-0463}{\usebox{\orcid} Patrick Laloyaux}}
    \And {\href{https://orcid.org/0000-0001-5302-6093}{\usebox{\orcid} Niels Bormann}}
    \And {\href{https://orcid.org/0000-0003-2808-0463}{\usebox{\orcid} Anthony McNally}}
    \And
    European Centre for Medium-Range Weather Forecasts (ECMWF)
}

\begin{document}

\maketitle

\begin{abstract}
Through a series of experiments, we provide evidence that the GraphDOP model - trained solely on meteorological observations, using no prior knowledge - develops internal representations of the Earth System state, structure and dynamics as well as the characteristics of different observing systems. Firstly, we demonstrate that the network constructs a unified latent representation of the Earth System state which is common across different observation types. For example, cloud structures maintain physical consistency whether viewed in predictions for satellite radiances from different sensors, or for direct in-situ measurements of the cloud fraction. Secondly, we show examples that suggest that the network learns to emulate viewing effects - learned observation operators that map from the unified state representation to observed properties. Microwave sounder limb effects and geometric viewing effects, such as sunglint in visible imagery, are both well captured. Finally, we demonstrate that the model develops rich internal representations of the structure of meteorological systems and their dynamics. For instance, when the network is only provided with observations from a single infrared instrument, it is able to infer unobserved, non-local structures such as jet streams, surface pressure patterns and warm and cold air masses associated with synoptic systems. This work provides insights into how neural networks trained solely on observations of the Earth System spontaneously develop coherent internal representations of the physical world in order to meet the training objective - enhancing our understanding and guiding future development of these models.
\end{abstract}

\section{Introduction}
\label{sec:intro}

Weather prediction is undergoing a transformation with data-driven models proving to be highly effective and in many ways surpassing the skill of physics-based counterparts (e.g. \cite{keisler2022}, \cite{bi2023nature}, \cite{lam2023}, \cite{lang2024}, \cite{Price2025GenCast}, \cite{Lang2024AIFSCRPS}). These models have primarily been trained on reanalysis datasets, notably ERA5 \citep{hersbach2020era5} produced by the European Centre for Medium-Range Weather Forecasts (ECMWF) under the Copernicus programme.

The analyses in these datasets are produced by data assimilation systems which combine observations with forecasts from numerical weather prediction (NWP) models that encapsulate our physical understanding of the Earth System \citep{rabier2000}. Gaps between sparse observations are filled using the physical relationships between variables and the structure of the meteorological systems defined by the physical models, and data-driven systems learn those relationships implicitly embedded in the data.

In contrast, the AI-Direct Observation Prediction (AI-DOP) approach, first proposed in \cite{mcnally2024}, seeks to learn directly from observational data. \cite{alexe2024} introduced GraphDOP - the first skilful end-to-end global medium-range weather prediction model trained exclusively from observations. Unlike other studies that rely on a combination of observational and analysis datasets \citep{Andrychowicz2023MetNet3,yuval2024,Allen2025Aardvark}, GraphDOP uses no prior physical knowledge; the relationships it learns come directly from observations of nature rather than from numerical modelling output. Given observations in a 12-hour input window, the network is given the training objective to predict the observations in the following 12-hour window.

A central assumption of this approach is that to be able to predict a future observation, the network must develop internal representations - a model - of the Earth System state and processes that the observation is sensitive to.

This principle is also true in other fields of machine learning. For example, in order to meet their training objective of next token prediction, large language models must learn representations of underlying concepts behind collections of words, as discussed e.g. in \cite{lindsey2025}. Some argue that the representations learned by different models converge towards an underlying "platonic" reality \citep{huh2024}. In the field of generative video modelling, researchers frequently use the concept of “world models” - internal representations of the physical world - which they argue emerge during training on observations of the world contained in the video frames (e.g., \cite{Ha2018,hafner2020,wang2024worlddreamer}). However, in the physical sciences there is still active debate about the extent to which data driven systems learn true physical relationships \citep{vafa2025_foundation_models,Bonavita2024_MLWeatherLimitations}.

The choice to train solely on observations is motivated not only by scientific curiosity but also by considerations of operational simplicity and sustainability in comparison to hybrid approaches. Output from NWP models and reanalyses has many strengths, providing spatial and temporal completeness and physically consistent fields. Networks trained on this output have already demonstrated significant skill but are also liable to inherit the limitations of those models, including their systematic errors. By excluding model-based fields, we avoid conflicting or contradictory signals that can arise in hybrid schemes when observations and reanalysis products are mixed in the same training dataset \citep{yuval2024}. Networks trained on observations have the potential to learn representations of physical processes that are observed but not fully represented in current NWP models or their analyses. This raises the possibility of extracting new information content from observational archives that is untapped by current data assimilation systems which could potentially allow both an improved model and enhanced initial conditions. However, questions remain about whether the observation coverage is complete enough to allow this approach to fulfill its potential.

With the aim of improving our understanding of the workings and behaviour of observation-driven end-to-end weather prediction models, this paper presents experiments exploring different aspects of the learned representations in GraphDOP. Section \ref{sec:unified_state} shows an example of a unified latent representation of the Earth System state. Evidence of learned representations of viewing effects are shown in Section \ref{sec:viewing_effects}. Section \ref{sec:struct_dynamics} examines the network's ability to infer unobserved structures via correlations with observed features - evidence of a representation of common meteorological structures, and their dynamics.

We hope that these insights will help guide future research and development in this exciting new field.

\section{Model and data description}
\label{sec:model_and_data}

\begin{figure}[htbp]
    \centering
    \includegraphics[width=0.95\textwidth]{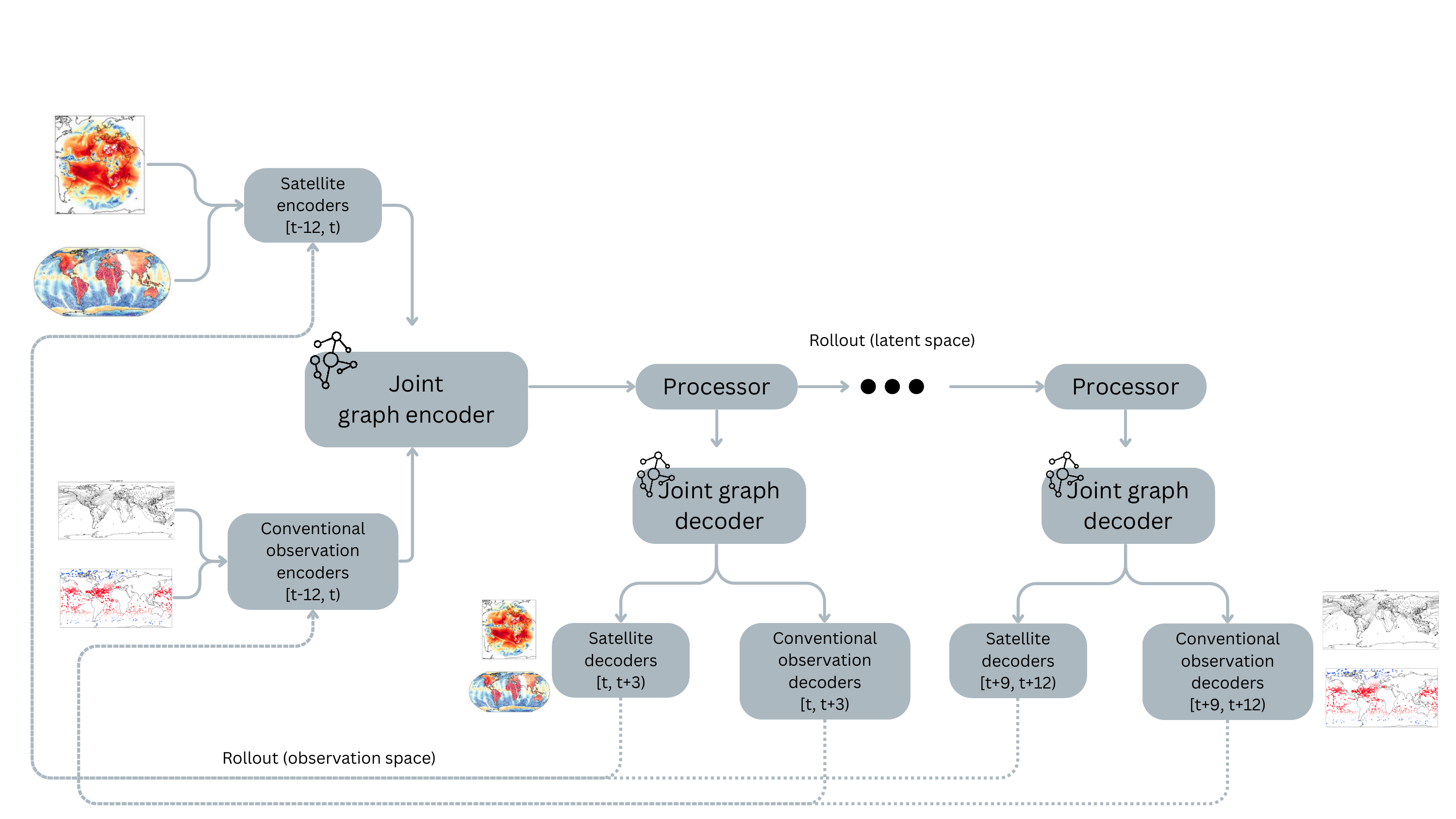}
    \caption{GraphDOP \citep{alexe2024} with 3-hour autoregressive time-stepping (rollout) in the latent space. During fine-tuning, the forecasted observations are fed back as inputs to produce the next forecast (rollout in observation space).} 
    \label{fig:dop_model_schematic}
\end{figure}

The experiments in this paper use the GraphDOP model described in \cite{alexe2024}. GraphDOP is following an encode-process-decode architecture where the input (output) observation data are jointly encoded (decoded) into (out of) a latent space representation. The encoder and decoder are attention based graph neural networks, while the processor is a transformer with a sliding attention window \citep{lang2024}. The processor evolves the latent representation throughout the forecast window. The GraphDOP model used in this paper encodes a single 12-hour window of observations; unlike the variant described by \cite{alexe2024}; however, it uses latent-space time stepping to decode a 12-hour window of target observations across four 3-hour process-decode steps, as illustrated in Figure \ref{fig:dop_model_schematic}. Forecasts produced using latent space time stepping were sharper and more accurate compared to the simpler alternative of decoding all observations inside the 12-hour target window in a single process-decode step. The latent representation is constructed on an o96 octahedral reduced Gaussian grid \citep{malardel2016}, at a spatial resolution of approximately 1$^o$, with 1024 features per grid node.

Since the observation locations and times change every day, the model must be able to make predictions for any arbitrary location and time within the output window. This means that during inference (forecasting) we are able to produce output on a regular grid.

The observation datasets used in this study differ from those in \cite{alexe2024}. Firstly, we use fewer observation types, concentrating mainly on the microwave satellite sounder instruments which have been found to be particularly well used in the current model. Unlike the previous study we also include data from cross-track microwave humidity sounders and Atmospheric Motion Vectors (AMVs). The full list of observation sources is provided in Table \ref{table:dop-observations} in the Appendix. In addition, we have made various improvements to the data quality control.

All experiments used a training dataset covering January 2013 to June 2022. The validation period was July to December 2022. Results are shown from the test period of January 2023.

\section{Unified representations of Earth System state}
\label{sec:unified_state}

\begin{figure}[htbp]
    \centering
    
    \begin{subfigure}[b]{\textwidth}
        \centering
        \begin{overpic}[width=0.75\textwidth]{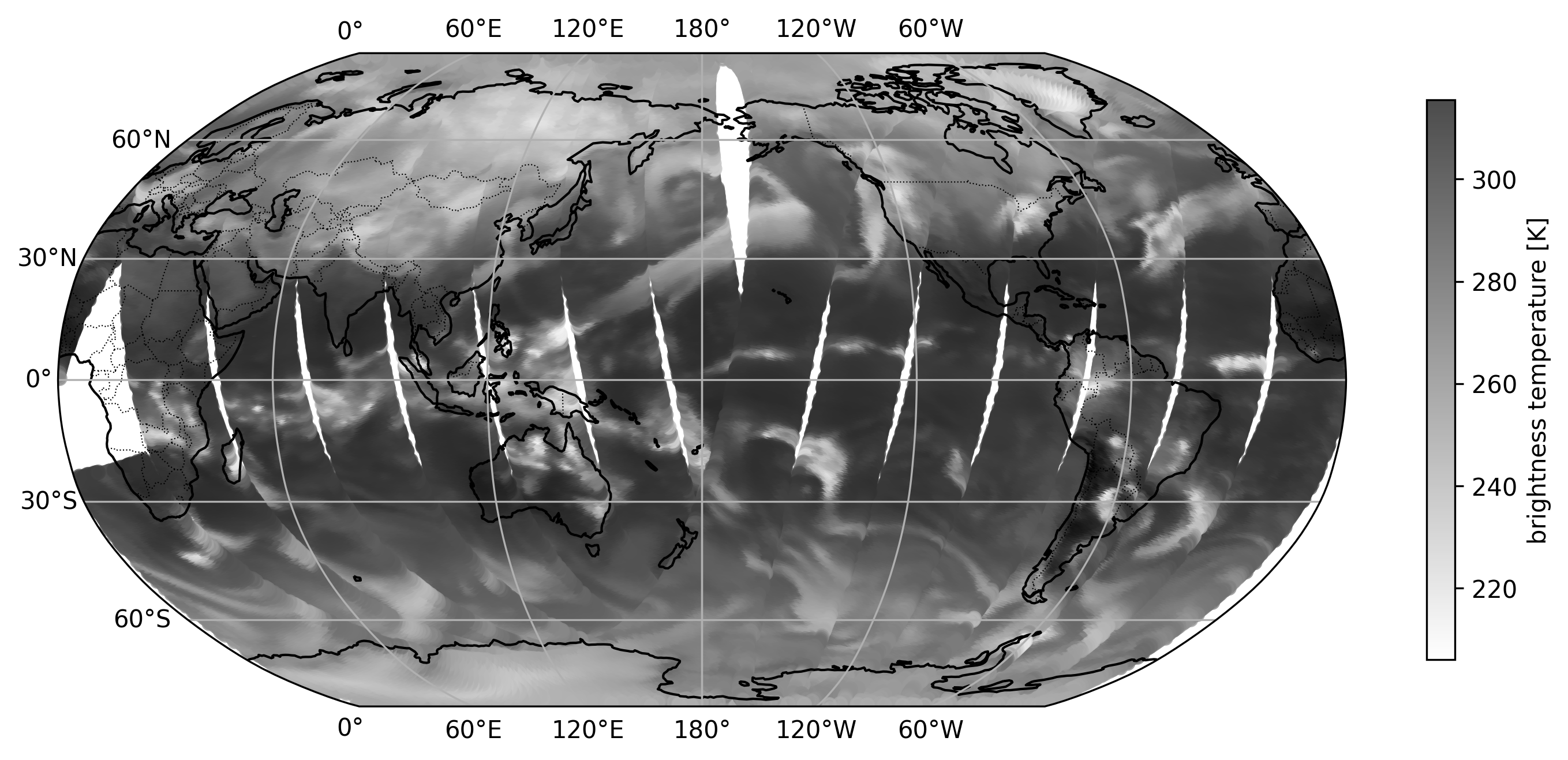}
            \thicklines
            \put(47,5){\color{yellow}\framebox(18,18){}}
        \end{overpic}
        \caption{Prediction for IASI channel 756 brightness temperature.}
        \label{fig1:iasi_ir}
    \end{subfigure}

    \vspace{0.5cm}

    \begin{subfigure}[b]{\textwidth}
        \centering
        \begin{overpic}[width=0.75\textwidth]{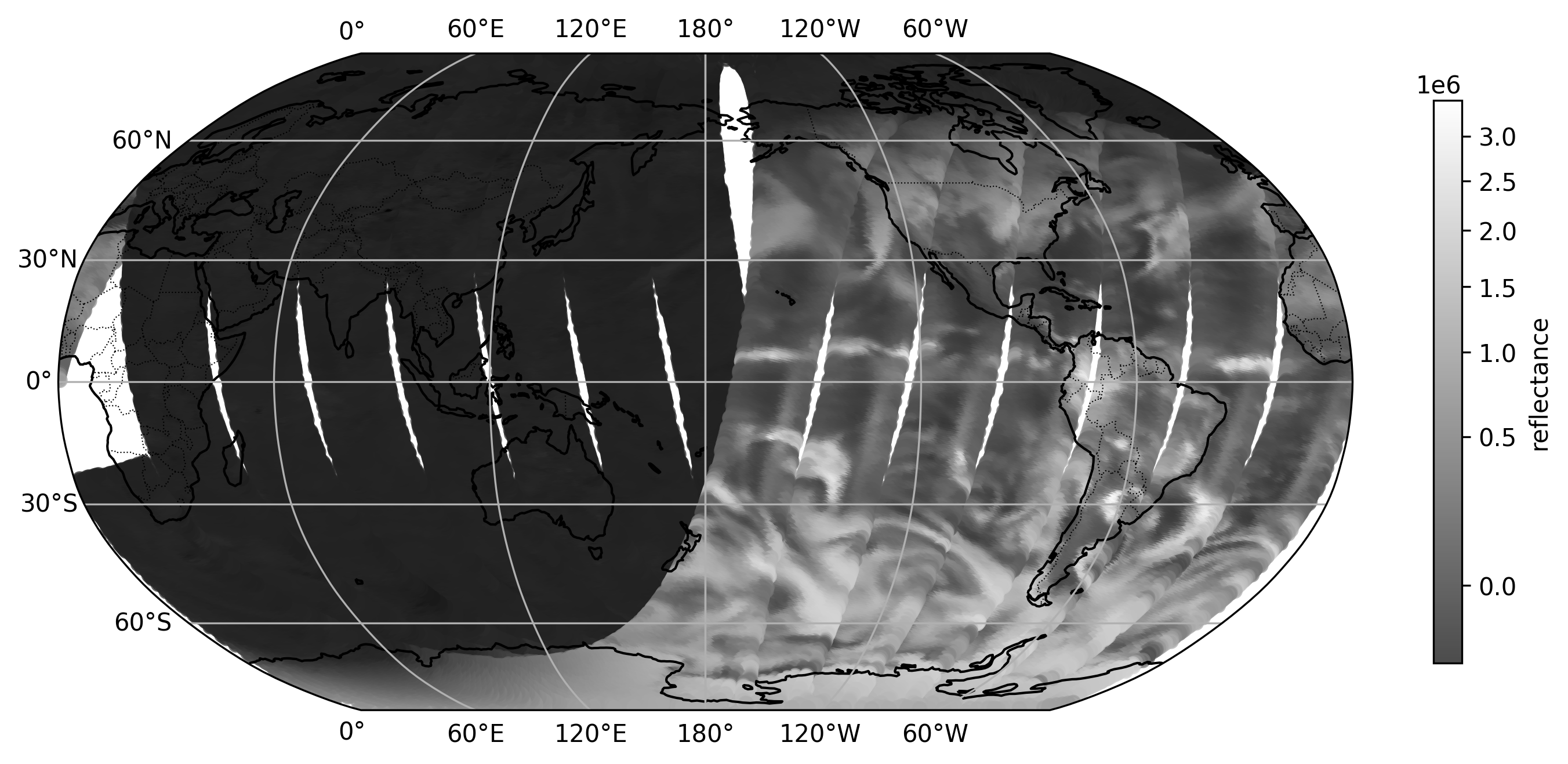}
            \thicklines
            \put(47,5){\color{yellow}\framebox(18,18){}}
        \end{overpic}
        \caption{Predicted AVHRR visible reflectance.}
        \label{fig1:avhrr_vis}
    \end{subfigure}

    \vspace{0.5cm}

    \begin{subfigure}[b]{\textwidth}
        \centering
        \begin{overpic}[width=0.75\textwidth]{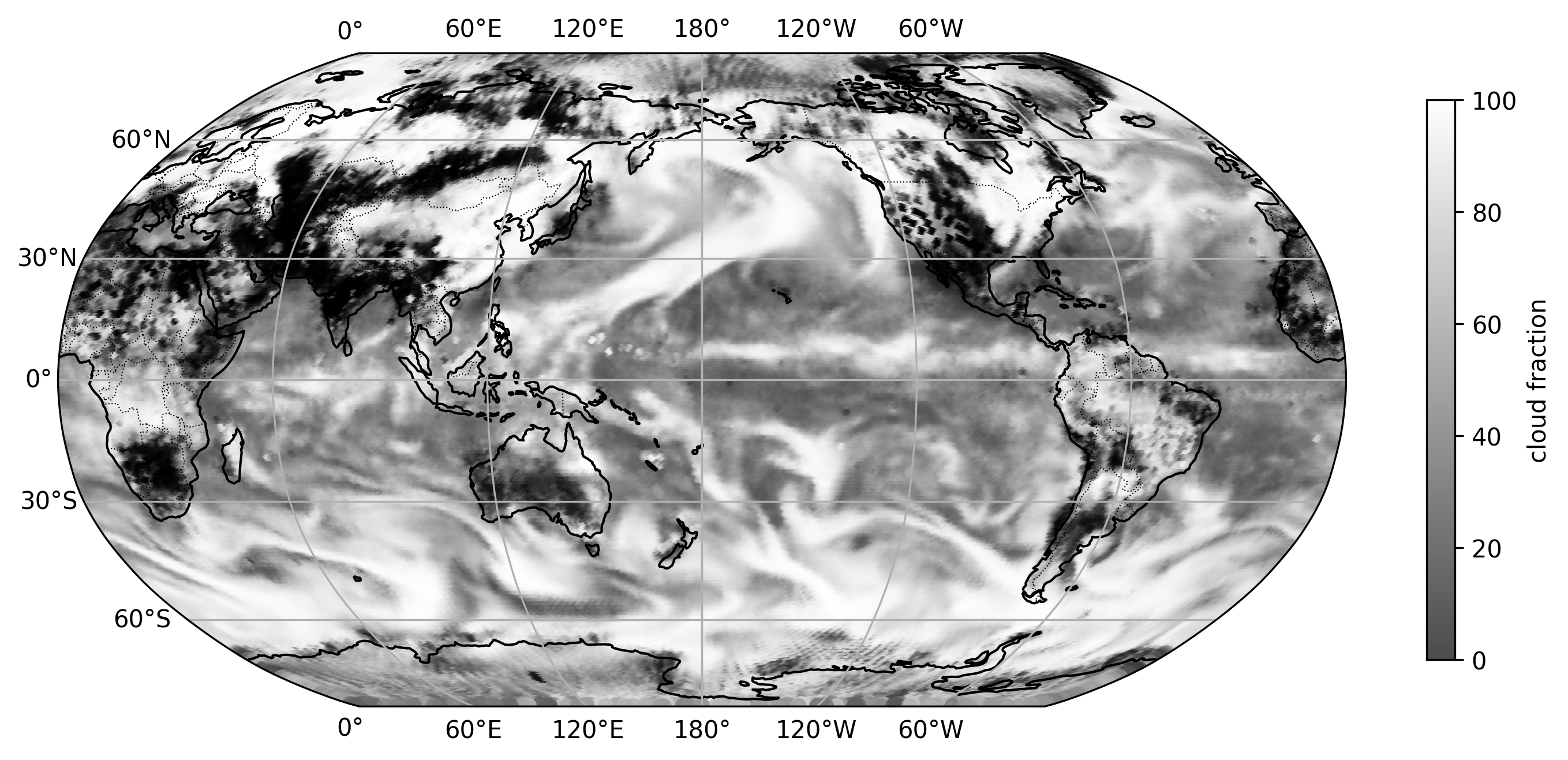}
            \thicklines
            \put(47,5){\color{yellow}\framebox(18,18){}}
        \end{overpic}
        \caption{Predicted cloud fraction from the SYNOP decoder.}
        \label{fig1:synop_cloud}
    \end{subfigure}

    \caption{Predictions in the first 12 h window from a forecast started
             12 January 2023 09 UTC for (a) IASI channel 756 (a window channel sensitive to clouds and the surface),
             (b) AVHRR visible reflectance and
             (c) SYNOP cloud fraction on an N320 grid.}
    \label{fig:cloud_panels}
\end{figure}

Being able to combine diverse information from a multitude of sensors, each sensitive to the underlying physical state in different ways, is a core requirement of any observation-driven model. Many meteorological observations are only \textit{indirectly} related to the physical state variables that we are interested in forecasting. For instance, the majority of the available information on upper-air temperature comes from satellite sounders which measure emitted radiation sensitive to the temperature and humidity integrated over deep layers of the atmosphere.

While the inputs and outputs to GraphDOP are purely in observation space, the encoders transform these observed variables into a shared latent space in the processor. In principle, this architecture allows GraphDOP to learn its own unified representation of the underlying physical state. In this section, we look for signs that this is occurring - using cloud as an example - by examining predictions for several observation types, each sensitive to cloud in different ways.

Firstly, SYNOP stations report fractional cloud cover as viewed from the surface. In the case of manual SYNOP stations, this is estimated by human observers. Since the clouds visible from a single observing site can be up to several hundred kilometres away, this observation is not a point value but representative of the cloud fraction across a region. For automatic SYNOP stations the cloud fraction can be inferred using ceilometers to measure the cloud across a short time window. Secondly, the IASI infrared sounder includes channels which are sensitive to clouds. For example, when clouds obscure the satellite's view of the surface, the brightness temperatures in the window channels are lower, corresponding to thermal emission from the cooler cloud tops. Finally, the AVHRR imager (on the same satellite as IASI) includes a visible channel which measures the reflectance of solar radiation. These images provide good information on cloud cover during daylight conditions thanks to the high albedo of the clouds.

Figure \ref{fig:cloud_panels} shows the predictions for each of these three observation types during the first 12 hour window of a forecast from a single case study. It is worth noting that other observation types, such as microwave instruments, also have sensitivity to clouds but are not considered in this example. The IASI and AVHRR predictions are made for the times and locations of the actual observations in the next window. The SYNOP cloud fraction predictions are made on an N320 grid (ca. 0.25-degree) mid-way through the window. Note that there will be a slight time mismatch (up to several hours) between the overpass of the satellite in the IASI/AVHRR predictions and the fixed prediction time used for the SYNOP cloud fraction.

The cloud structures show good spatial consistency across all three observation types, for example as highlighted in the yellow box. It is particularly noteworthy that the SYNOP cloud fraction decoder makes fine-grained forecasts of cloud structures in the south Pacific; an area which rarely has any SYNOP cloud reports. This indicates that the network has correctly identified that these very different observation types are observing the same underlying physical structure in the atmosphere (clouds) and has developed its own internal notion of cloud in the latent state which the satellite brightness temperature decoders and the SYNOP cloud decoders both map from. This relationship generalises well to new locations which have never seen SYNOP reports before.

Some artifacts can be seen in the SYNOP cloud cover forecast over some land areas e.g. the patches with low cloud fraction over the western United States and North Africa. Although the cause of these features is still under investigation, one hypothesis is that they are artifacts related to individual SYNOP stations, with shapes aligned to the internal processor grid.

The AVHRR visible channel predictions are notable since the areas that are sunlit during the output window, would have been in the Earth's shadow during the input window 12-hours earlier. Since the network would not have been able to use the visible imagery at those nighttime locations to predict the cloud in the next window the information must have come from other cloud-sensitive sensors which informed the network's internal representation of cloud.

The relationship between infrared brightness temperature and cloud is highly situation dependent. The model is able to discriminate well between the cold surfaces of Siberia in the IASI window channel and cloud features. It is also apparent from the predicted infrared imagery that the network has a representation of the height of the cloud top correctly predicting differences in cloud top temperature.

\section{Representations of viewing effects}
\label{sec:viewing_effects}

To be able to extract information on the state of the Earth System from observations, the network needs a good representation of the characteristics of the observing systems themselves. In this section we show two examples of viewing effects which the network has successfully learned to emulate.

\subsection{Limb effects}
\label{subsec:limb}

Impact studies consistently show that microwave sounders are the single largest contributor to forecast skill in current NWP systems \citep{Bormann_2019_ECMWF_TM839,Samrat_2025_QJRMS}. These sounding instruments measure the upwelling microwave radiance arising from thermal emission across a deep layer of the atmosphere with each channel having a weighting function which peaks at a different height in the atmosphere.

One of the most common types of instruments used for this purpose are the cross-track microwave sounders, such as ATMS and AMSU-A. These scan from side to side perpendicular to the track of the satellite along its orbit. The observed brightness temperature depends strongly on the viewing zenith angle. When viewing at nadir, the atmospheric path length is shorter and extends deeper into the atmosphere and so the brightness temperatures are representative of temperatures closer to the surface. In the presence of a temperature gradient in the vertical, the brightness temperatures vary between the nadir and the edges of the swath. These are known as limb effects.

The magnitude of these limb effects can be significantly larger than the meteorological signals that we wish to extract. Therefore, to be able to extract atmospheric temperature and humidity signals from microwave sounder data (to inform the network’s latent representation of these physical quantities), accurate modelling of limb effects is crucial.

\begin{figure}
  \centering
  \begin{subfigure}[b]{0.48\textwidth}
    \centering
    \includegraphics[width=\textwidth]{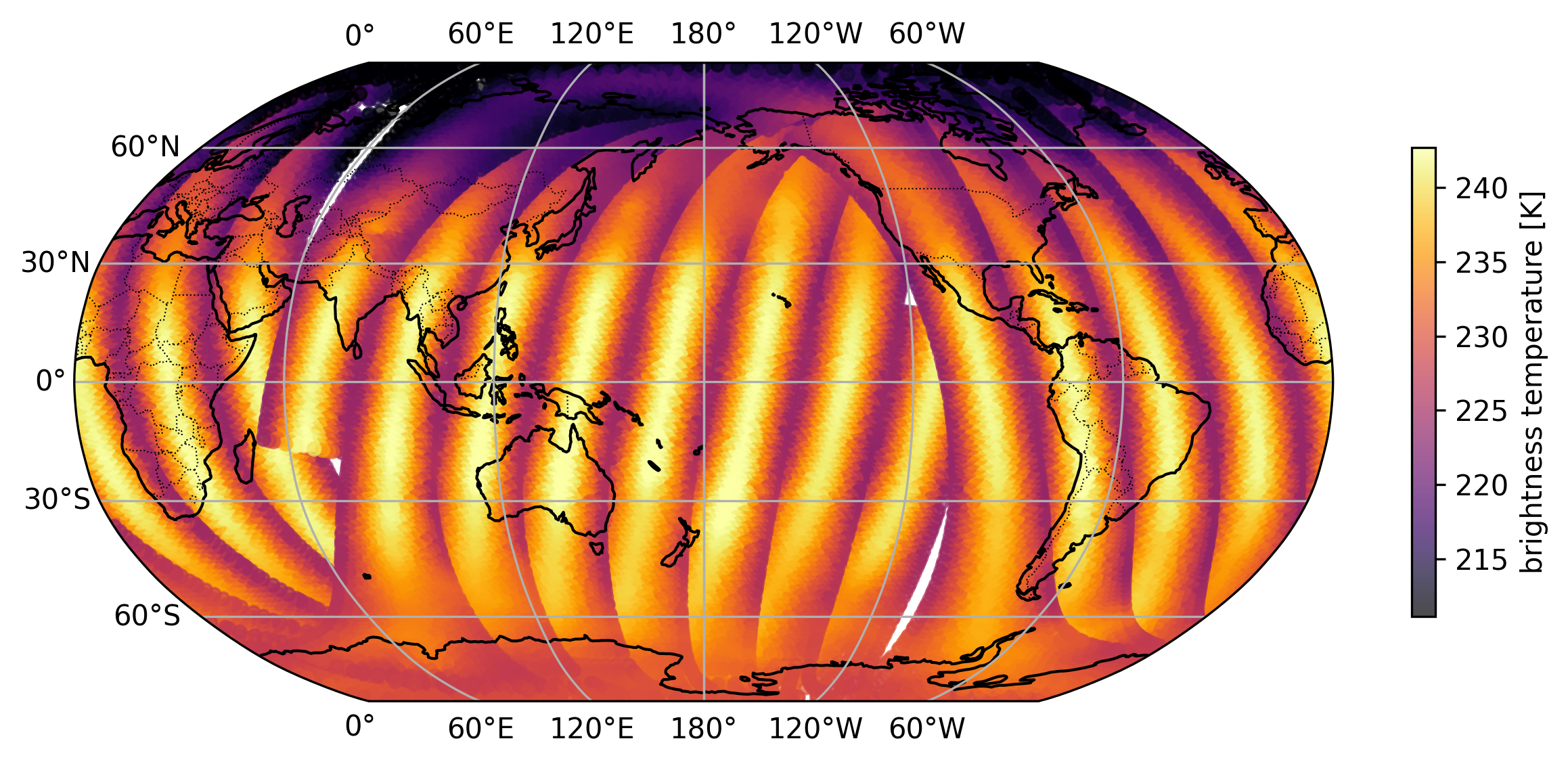}
    \caption{Prediction}
    \label{fig:limb_pred}
  \end{subfigure}
  \hfill
  \begin{subfigure}[b]{0.48\textwidth}
    \centering
    \includegraphics[width=\textwidth]{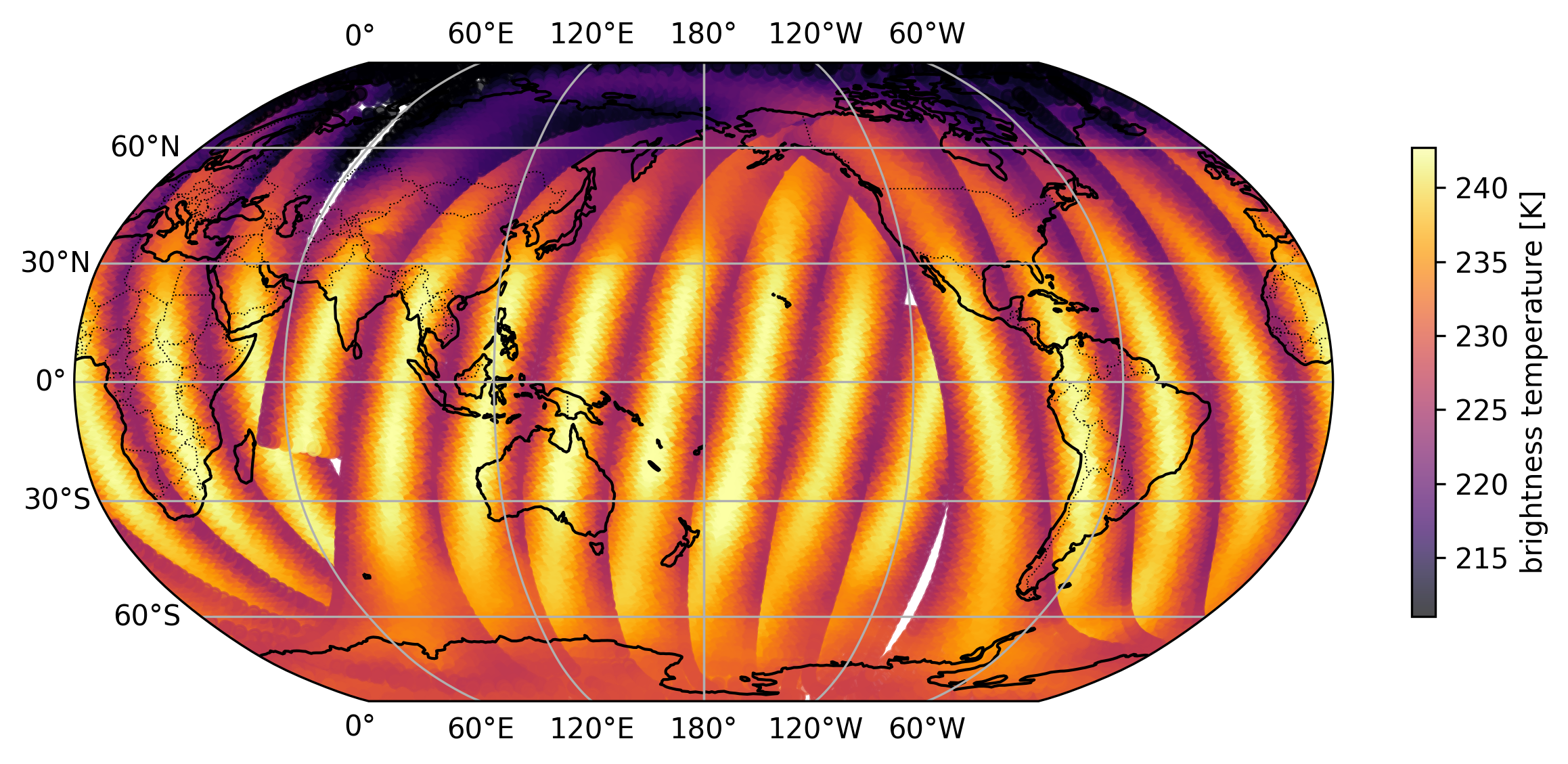}
    \caption{Target}
    \label{fig:limb_target}
  \end{subfigure}

  \vspace{1em}
  \begin{subfigure}[b]{0.55\textwidth}  
    \centering
    \includegraphics[width=\textwidth]{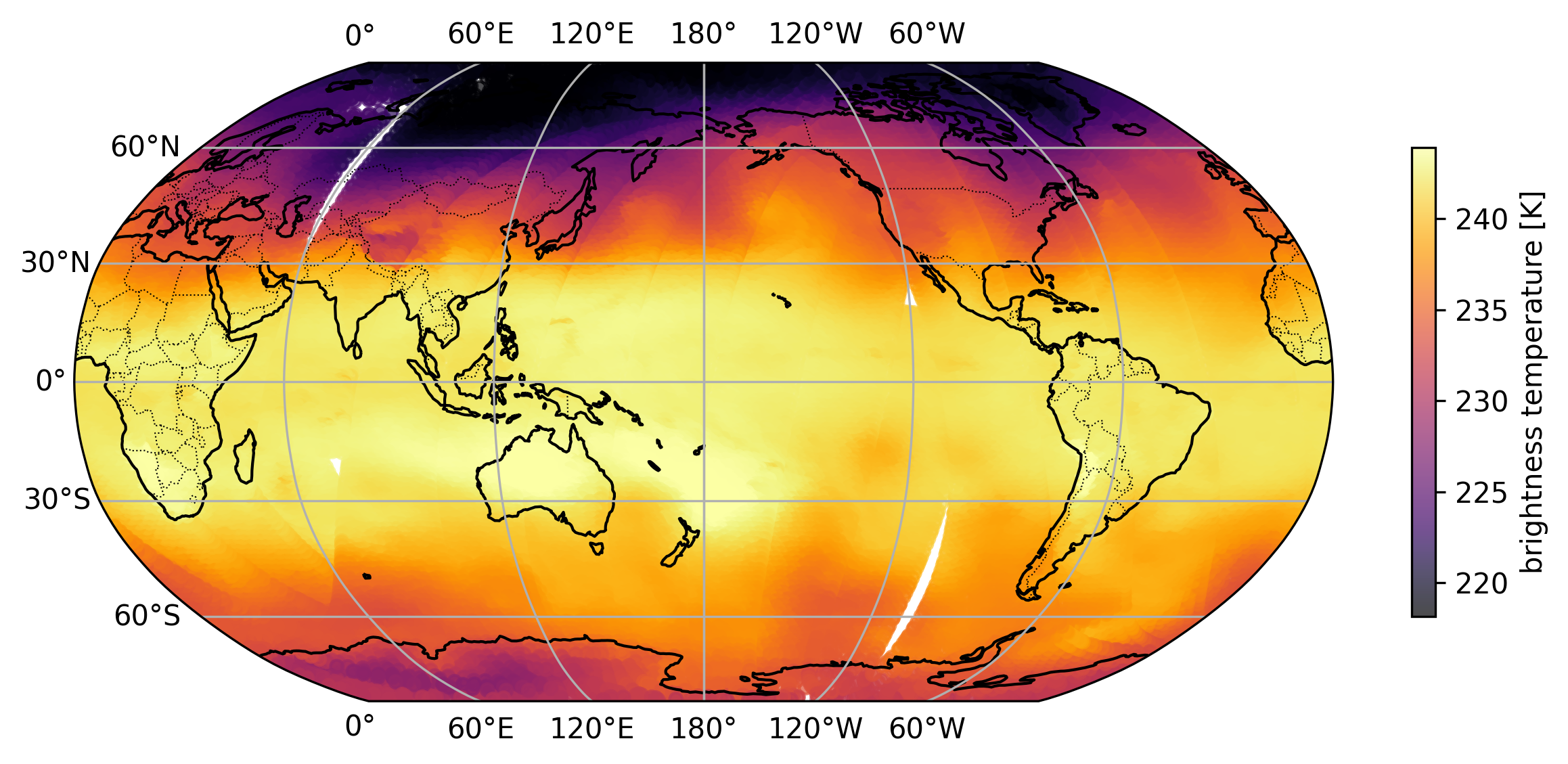}
    \caption{Predictions with viewing‐zenith angle fixed at 0°}
    \label{fig:limb_zero}
  \end{subfigure}

  \caption{(a) Predicted and (b) target brightness temperatures for NPP ATMS channel 7 during the first prediction window (10 Jan 2023 21 UTC – 11 Jan 2023 09 UTC). Bottom row: same as (a) but with the viewing-zenith angle set to zero for all fields of view.}
  \label{fig:limb_combined}
\end{figure}

\begin{figure}
    \centering
    \includegraphics[width=\textwidth]{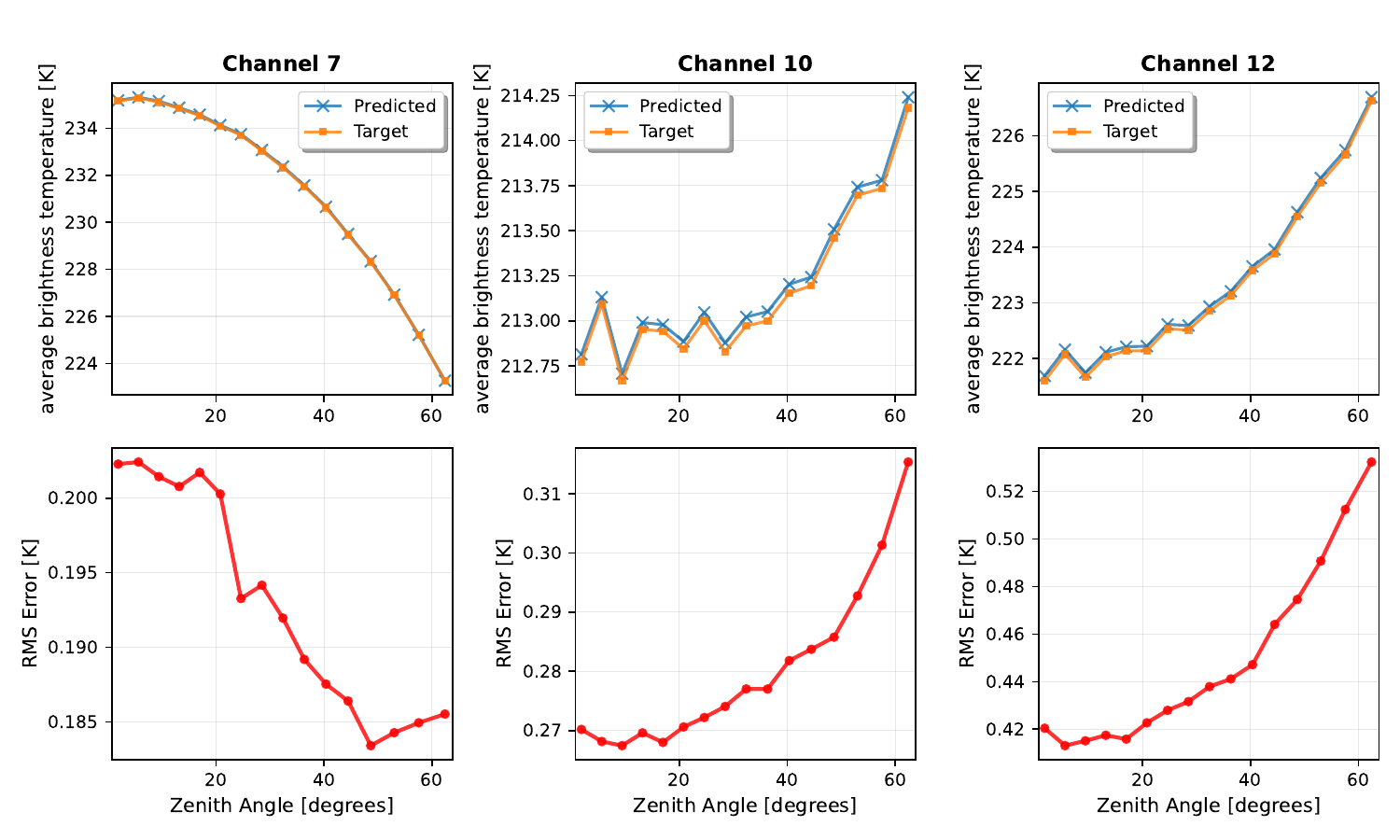}
    \caption{Average brightness temperature as a function of zenith angle (top row) in the predictions and the targets, for NPP ATMS channels 7 (left), 10 (centre) and 12 (right). The RMS error in the predictions is shown on the bottom row. All statistics are calculated in the first 12-hour prediction window, aggregated across 24 forecasts in January 2023.}
    \label{fig:limb_rms}
\end{figure}

Alongside information on the latitude, longitude and time, the network is provided with the viewing zenith angle for which the prediction is to be made. Figure \ref{fig:limb_combined} shows the predicted and observed brightness temperatures for ATMS channel 7 (a temperature sounding channel peaking in the upper troposphere). The strong limb effects are clearly visible in both the observed and predicted values.

The root mean squared error as a function of viewing zenith angle is shown in Figure \ref{fig:limb_rms}. Some residual error as a function of zenith angle remains, indicating that the limb effects are not modelled perfectly or that other aspects are not fully captured, such as the horizontal displacements arising from slanted viewing paths \citep{Bormann2017}. However, the magnitude of the difference in RMS error between nadir and the edge of the scan is small compared to the magnitude of the limb effect. For example, for channel 7 the RMS error is around 0.2K at nadir and around 0.185K at the limb, i.e. a 0.015K difference in RMS error compared to a limb effect on the order of 10K (as can be seen from the plots in the top panels).

Figure \ref{fig:limb_zero} shows the predicted brightness temperatures for ATMS channel 7 in an experiment where the zenith angle was artificially set to zero for all fields of view in the output. 

The limb effects are no longer visible and the variations in brightness temperature associated with different air masses become apparent. This demonstrates that the network has correctly learned the relationship between the zenith angle and the limb effects, which generalises to situations not seen during the training.

Whether the network has simply learned to apply a mean correction as a function of zenith angle, or is doing a more sophisticated modelling involving the vertical temperature gradients with respect to the changing weighting functions will be the subject of further study.

\subsection{Sunglint}
\label{subsec:sunglint}

\begin{figure}
    \centering
    \begin{subfigure}[b]{0.48\textwidth}
        \centering
        \includegraphics[width=\textwidth]{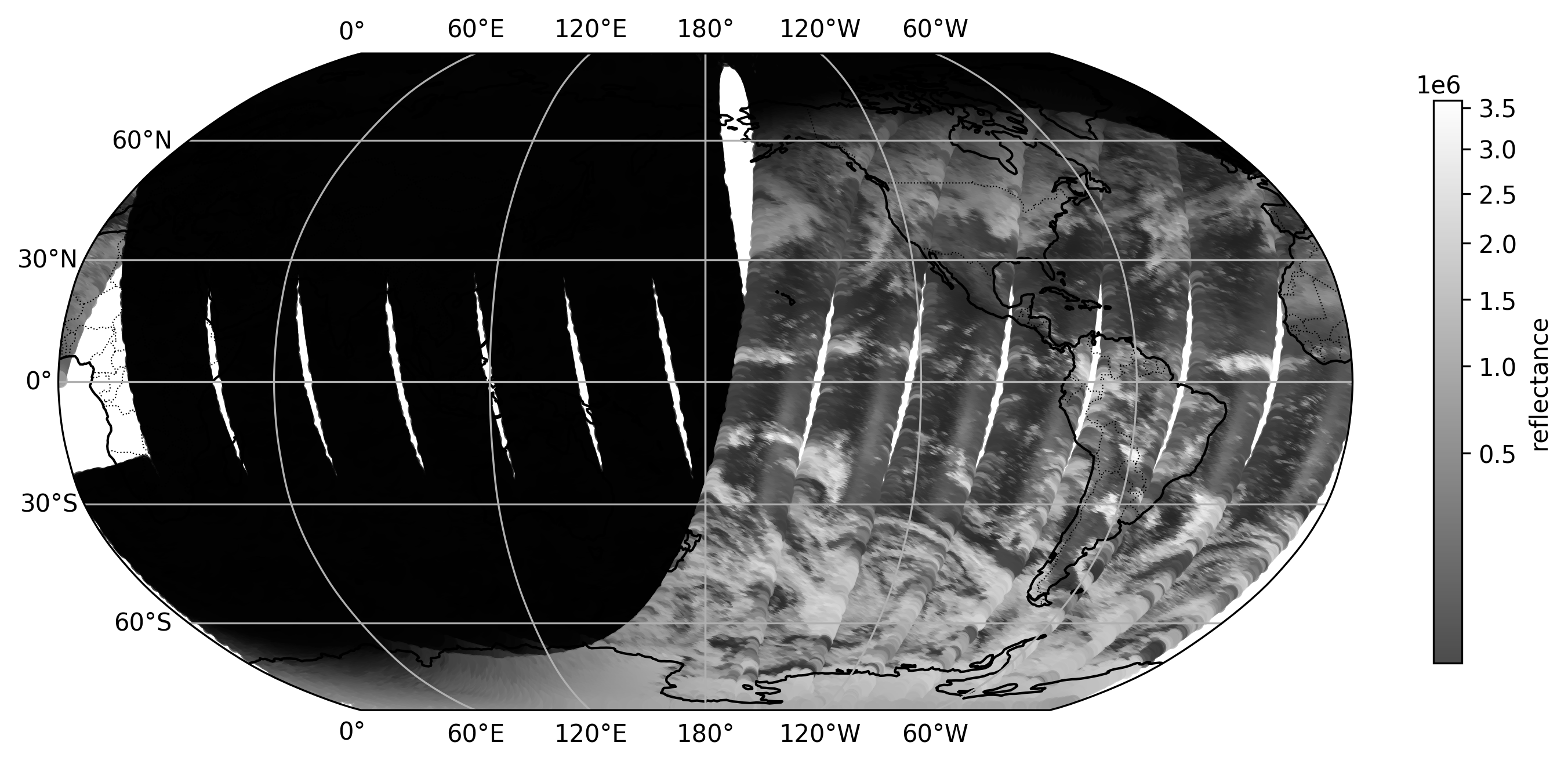}
        \caption{Target}
        \label{fig:vis_target}
    \end{subfigure}
    \hfill 
    \begin{subfigure}[b]{0.48\textwidth}
        \centering
        \includegraphics[width=\textwidth]{figs/vis_fig1.png}
        \caption{Prediction}
        \label{fig:vis_pred}
    \end{subfigure}

    \caption{Observed (left) and predicted (right) AVHRR visible reflectance in the first 12-hour window from a forecast from January 12th, 2023 at 09 UTC. Sunglint from the ocean surface is visible on the eastern half of each swath in the equatorial regions.}
    \label{fig:vis}
\end{figure}

The reflection of sunlight from ocean surfaces (known as sunglint) is observed in certain channels of satellite imager instruments. For example, in the AVHRR visible channel (see Figure \ref{fig:vis_target}) sunglint can be seen as diffuse stripes of higher reflectance along the track of the orbit, peaking in the equatorial regions. The increased reflectance associated with the sunglint is not a property of the state of the surface being observed but rather a viewing effect dependent on the viewing geometry. 

The prediction of the visible channel from GraphDOP is shown in Figure \ref{fig:vis_pred}. It can be seen that the network accurately models the presence of sunglint, with the sunglint peaking at the same latitude as that of the observations. This peak latitude is found to vary realistically with season (not shown).

Visible channels can be a useful observation to detect the presence of clouds, particularly those with low cloud tops which may be harder to discern from infrared channels since the cloud top temperature is similar to that of the surface. However, sunglint has the potential to contaminate this signal as the reflectance of sunlight from the ocean surface can be of the same magnitude as that of the reflectance from clouds. By comparing the sunglint predictions against the gridded SYNOP cloud fraction plot in Figure \ref{fig1:synop_cloud}, it can be seen that the model has correctly determined that these stripes of increased reflectance do not map onto its internal representation of cloud.

\section{Representations of meteorological structures and their dynamics}
\label{sec:struct_dynamics}

In data assimilation systems, the physical model plays two important roles:
\begin{enumerate}[label=(\roman*)]
  \item to ensure physically consistent relationships both spatially and between variables and, 
  \item to temporally propagate information from observations throughout the assimilation window.
\end{enumerate}

In this section, we look for signs that GraphDOP's learned representations of Earth System structures and their dynamics are able to perform a similar role.

\begin{figure}[ht]
    \centering
    \begin{subfigure}[b]{0.48\textwidth}
        \centering
        \includegraphics[width=\textwidth]{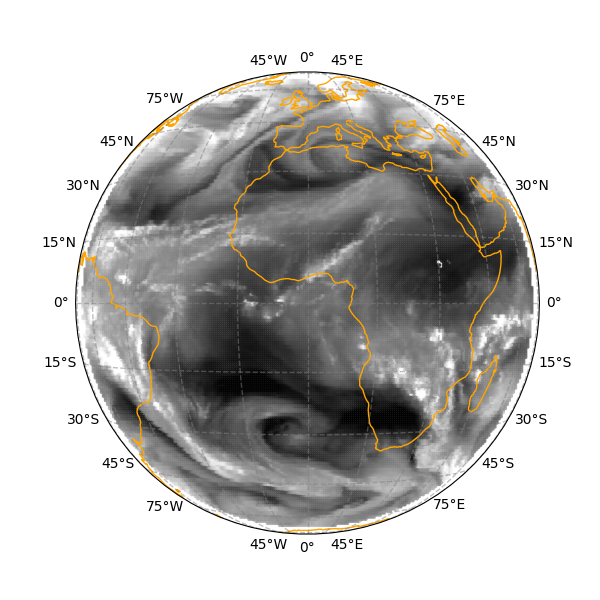}
        \caption{channel 5}
        \label{fig:seviri_ch5}
    \end{subfigure}
    \hfill 
    \begin{subfigure}[b]{0.48\textwidth}
        \centering
        \includegraphics[width=\textwidth]{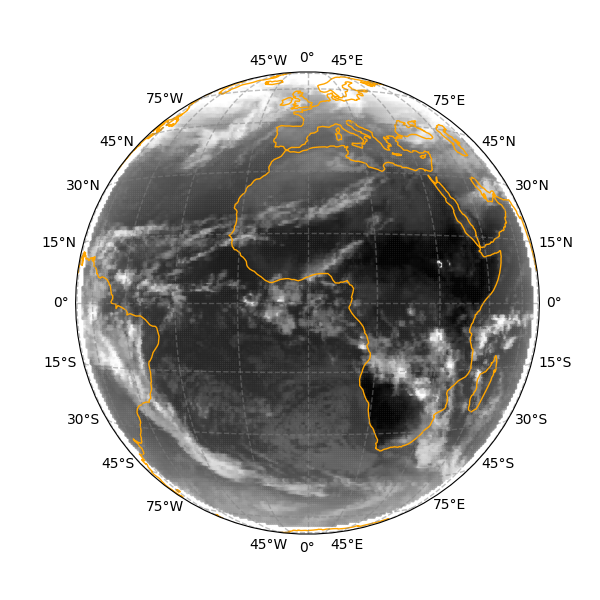}
        \caption{channel 8}
        \label{fig:seviri_ch8}
    \end{subfigure}

    \caption{Observed SEVIRI channel 5 (left) and channel 8 (right) at 08:45 UTC on Jan 6th, 2023, at the end of the input window.}
    \label{fig:seviri}
\end{figure}

\begin{figure}[H] 
    \centering

    \begin{subfigure}[b]{\textwidth}
        \centering
        \includegraphics[width=0.8\textwidth]{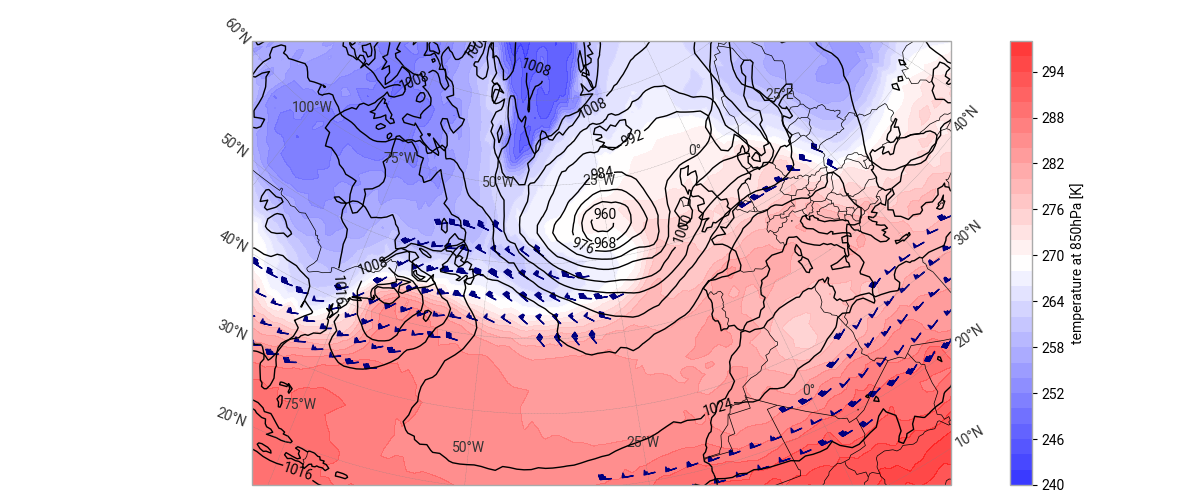}
        \caption{GraphDOP prediction from an experiment using only SEVIRI inputs}
        \label{fig:geos_predict_zoom}
    \end{subfigure}

    \vspace{0.25cm} 

    \begin{subfigure}[b]{\textwidth}
        \centering
        \includegraphics[width=0.8\textwidth]{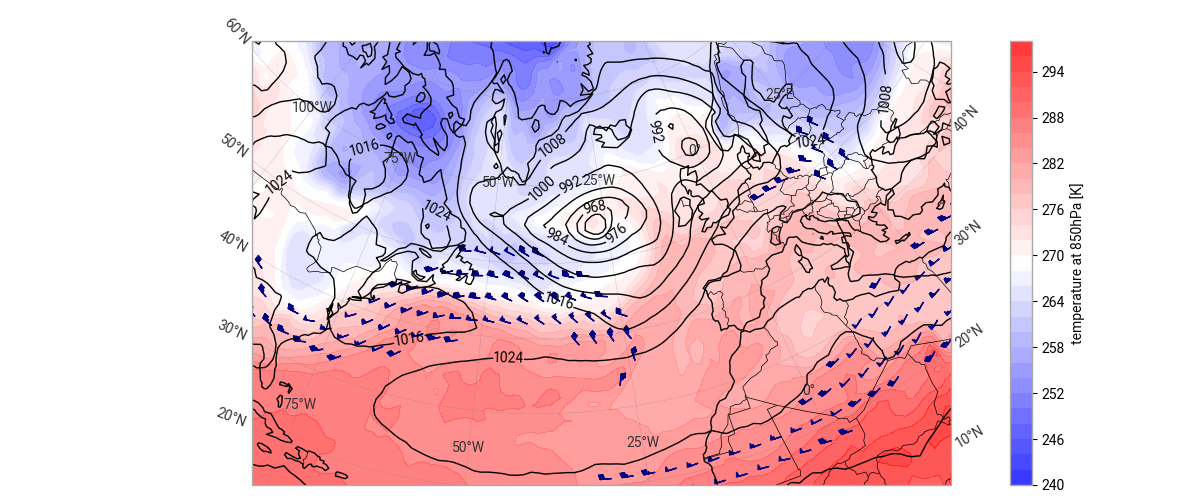}
        \caption{ECMWF operational analysis}
        \label{fig:geos_analysis_zoom}
    \end{subfigure}
   
    \caption{Mean sea level pressure in hPa (contours), temperature at 850~hPa (shaded) and wind speed at 200~hPa (barbs mark winds above $40~\mathrm{m\,s^{-1}}$) for the GraphDOP prediction (upper panel) and operational analysis (lower panel) on January 6th, 2023 at 12 UTC in an experiment which was only given SEVIRI data as input.}
    \label{fig:inferred_structures_zooom}
\end{figure}

\subsection{Structure of meteorological systems}
\label{subsec:structure}

\subsubsection{Inferring structures that are not directly observed}

To investigate GraphDOP's ability to infer unobserved and non-local meteorological structures, a separate training run was conducted where the only input was SEVIRI brightness temperatures from Meteosat-10 and 11. The training target was to predict the SEVIRI brightness temperatures as well as a range of conventional and Atmospheric Motion Vector observations globally in the following 12-hour window. Full details of the experiment setup are provided in Appendix \ref{appendix:geos_exp_setup}.

Since the Meteosat satellites are positioned in geostationary orbit above the equator at 0 degrees longitude, they provide continuous observations of the same hemisphere, with no coverage outside of this area. SEVIRI measures top-of-atmosphere radiances in multiple spectral bands spanning the visible, near-infrared, and thermal infrared. It includes channels at the water-vapour absorption bands (\(6.2\,\mu\mathrm{m}\) and \(7.3\,\mu\mathrm{m}\)) most sensitive to upper and mid-tropospheric humidity, as well as infrared window bands (e.g. \(10.8\,\mu\mathrm{m}\) and \(12.0\,\mu\mathrm{m}\) ) primarily sensing thermal emission from clouds and the surface. The visible channels are not currently included in the GraphDOP training dataset.

An example of the input observations from two of the eight channels at the end of the input window is shown in Figure \ref{fig:seviri}. Figure \ref{fig:geos_predict_zoom} shows examples of gridded output for several meteorological fields from the conventional observation decoders, three hours into the output window. This can be compared against the operational 4D-Var analysis from the same time in Figure \ref{fig:geos_analysis_zoom}.

It can be seen that GraphDOP was able to infer the location and depth of the extra-tropical cyclone to the south of Iceland quite accurately despite the input observations having no direct sensitivity to surface pressure. It correctly infers the warm and cold air masses wrapping around the centre of the low pressure (as indicated by the temperature at 850~hPa). The intensity and location of the jet stream (as shown by the 200~hPa wind speed) are also well captured.

Although the channels in SEVIRI have no direct sensitivity to temperature at 850~hPa and mean sea level pressure, the network is able to indirectly infer these based on learned relationships with properties that are observed. This is evidence consistent with GraphDOP having learned the spatial structure of synoptic systems.

In the case of the upper-level winds, the mechanism by which these are inferred is unclear. One possibility is that the network is able to infer them via feature tracking of observed upper tropospheric features in successive frames of the SEVIRI imagery in the same way that Atmospheric Motion Vectors are derived. Alternatively, the network may be able to infer the winds from a single frame by learning the correlations of the winds with spatial patterns in the imagery. Further work would be needed to better understand the mechanism used by the network in this case.

\subsubsection{Inferring non-local atmospheric structures in regions with no observations}

\begin{figure}[htbp]
    \centering
    \begin{subfigure}[b]{\textwidth}
        \centering
        \includegraphics[width=0.8\textwidth]{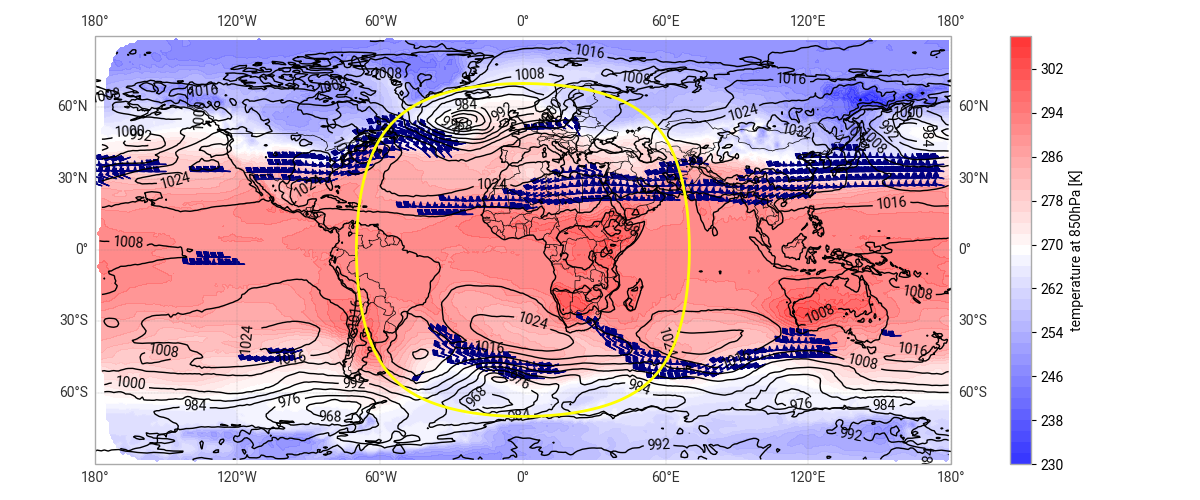}
        \caption{GraphDOP prediction from experiment using only SEVIRI inputs}
        \label{fig:geos_predict}
    \end{subfigure}

    \vspace{0.5cm} 

    \begin{subfigure}[b]{\textwidth}
        \centering
        \includegraphics[width=0.8\textwidth]{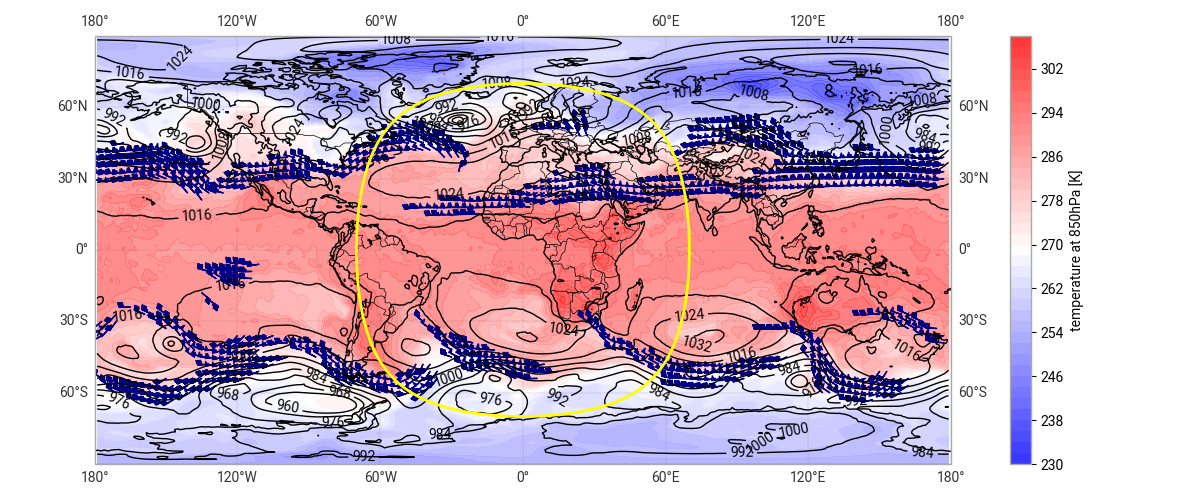}
        \caption{Operational Analysis}
        \label{fig:geos_analysis}
    \end{subfigure}    
    
   \caption{Same as Figure \ref{fig:inferred_structures_zooom} but showing data globally. The extent of SEVIRI coverage is highlighted in yellow.}    \label{fig:inferred_structures}
\end{figure}

Further understanding of GraphDOP's internal representations can be revealed by looking at the global predictions in this experiment (see Figure \ref{fig:inferred_structures}). Since the only inputs provided in this experiment came from SEVIRI which only sees one side of the Earth, much of the planet had no incoming observations. The locations of synoptic systems and jet streams are well captured on the hemisphere observed by SEVIRI (centred on 0 degrees longitude). However, outside this region the GraphDOP predictions (Figure \ref{fig:geos_predict}) become more zonal and featureless. In the absence of any input observations the model appears to relax towards a climatological state which it has learned in order to predict the training targets (which were still present globally). For example, the deep low pressure systems in the North Pacific are missing completely, as is the branch of the jet stream to the north of the Himalayas.

\begin{figure}
    \centering
    \includegraphics[width=0.8\textwidth]{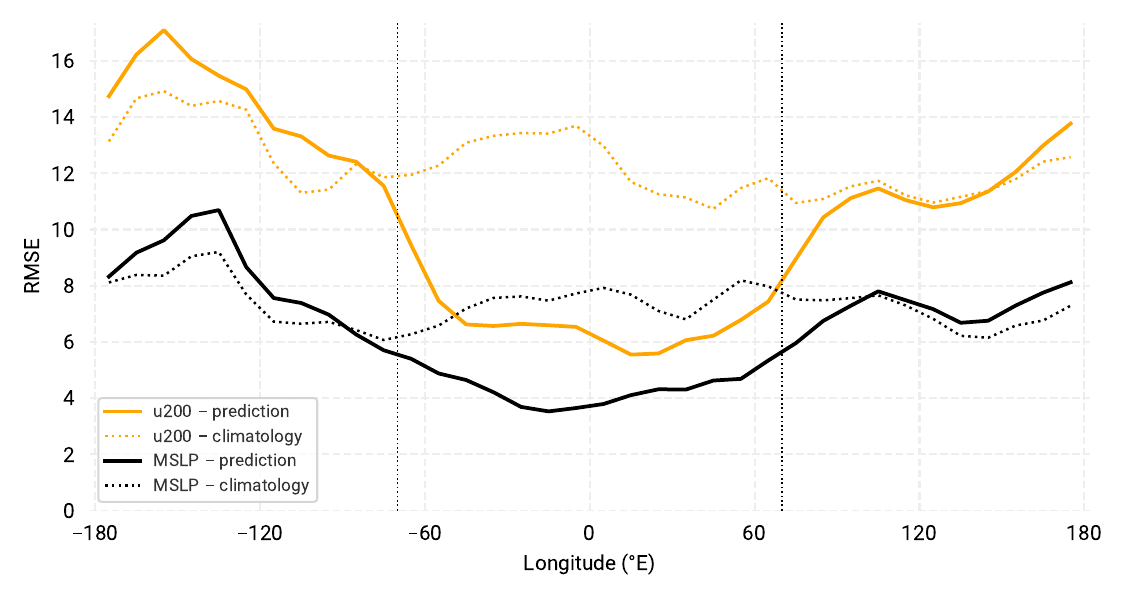}
    \caption{Area-weighted RMS difference between GraphDOP t+12h predictions and the ECMWF operational analyses for mean sea level pressure [hPa] (solid black) and the u-component of the wind at 200~hPa [ms$^{-1}$] (solid orange) as a function of longitude in 10-degree bins. Results are aggregated across all GraphDOP forecasts in January 2023. Climatological baselines derived from ERA5 are also shown (dotted lines). The extent of the region observed by SEVIRI is marked by vertical dotted lines.}
    \label{fig:rms_by_lon}
\end{figure}

Figure \ref{fig:rms_by_lon} shows the RMS error for upper-level winds and for mean sea level pressure as a function of longitude. A climatological baseline derived from ERA5 is also provided for each variable. We can see that the prediction errors are substantially lower at the longitudes observed by SEVIRI. Outside of the observed region the errors increase to be broadly in line with the climatological errors. We see some evidence that there are skilful predictions slightly outside of the observed region, for example between 70 and 100 degrees east. This is evidence that the network is able to infer non-local atmospheric structure based on correlations with observed features.

This is consistent with the downstream flow of information from observations at the beginning of the input window to the forecast time. It might be expected that with the use of a longer input window the information from earlier observations would continue to propagate downstream and extend the area of skilful predictions further outside of the SEVIRI domain. Similarly, this could also be achieved by cycling information from earlier windows through a background state. This will be the subject of further research.

\begin{figure}[H]
    \centering
    \begin{minipage}{0.45\textwidth}
        \centering
        \includegraphics[width=\linewidth]{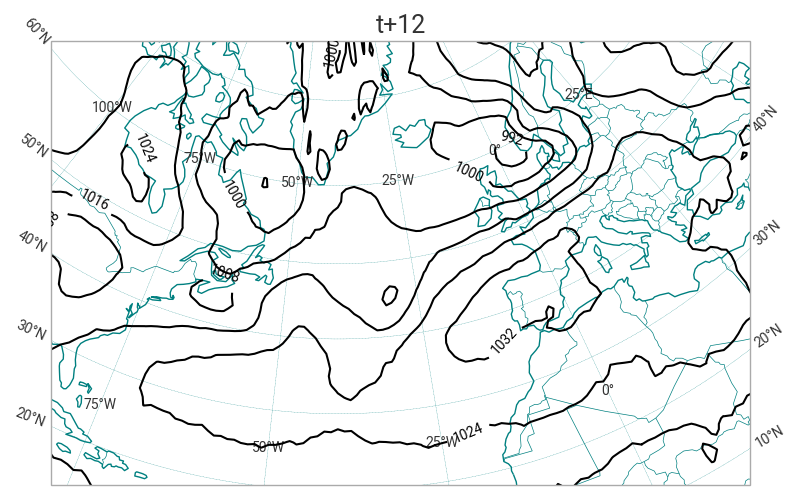}
    \end{minipage}
    \hfill
    \begin{minipage}{0.45\textwidth}
        \centering
        \includegraphics[width=\linewidth]{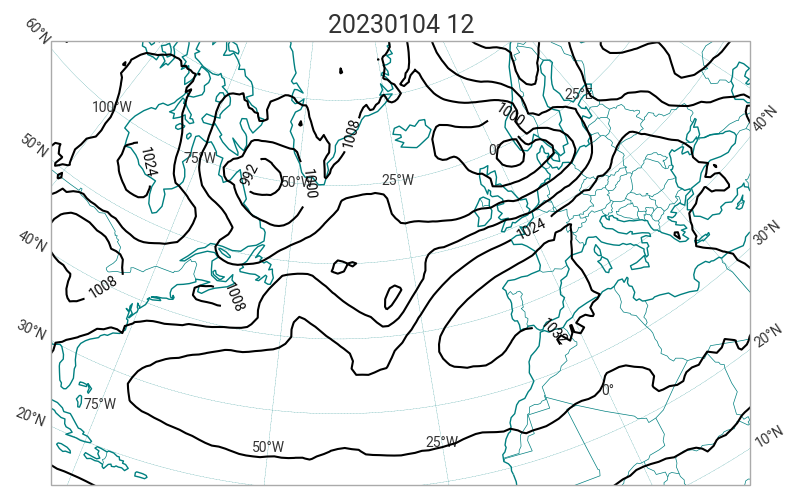}
    \end{minipage}
    
    \vspace{0.5cm}
    
    \begin{minipage}{0.45\textwidth}
        \centering
        \includegraphics[width=\linewidth]{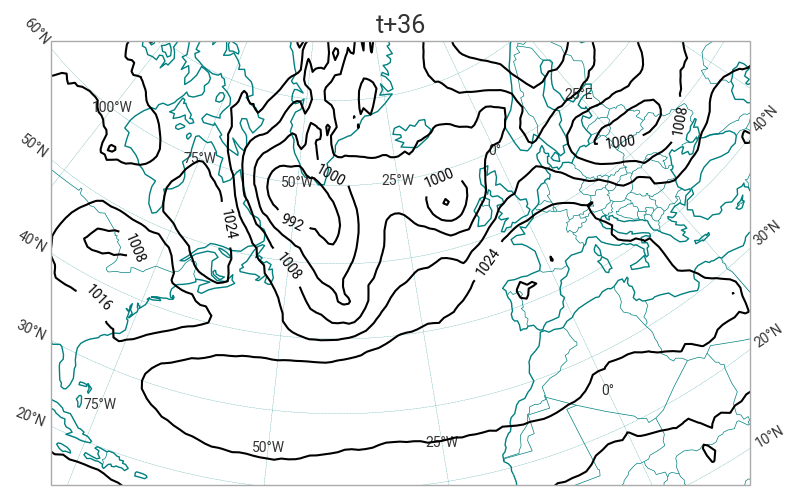}
    \end{minipage}
    \hfill
    \begin{minipage}{0.45\textwidth}
        \centering
        \includegraphics[width=\linewidth]{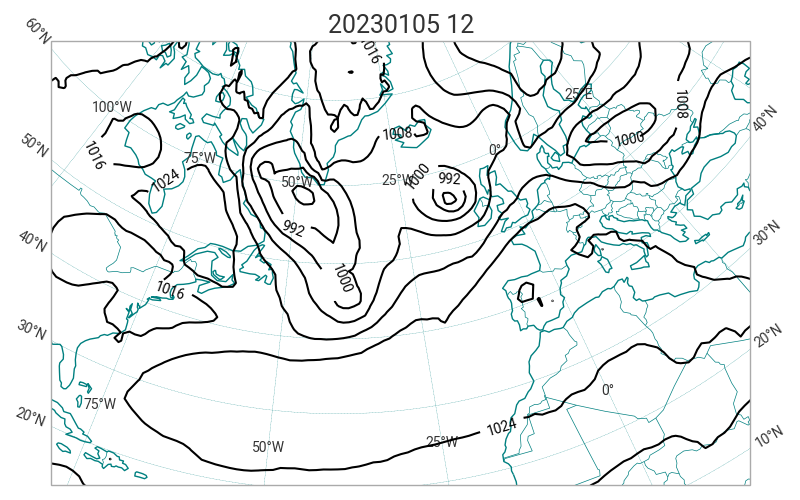}
    \end{minipage}
    
    \vspace{0.5cm}
    
    \begin{minipage}{0.45\textwidth}
        \centering
        \includegraphics[width=\linewidth]{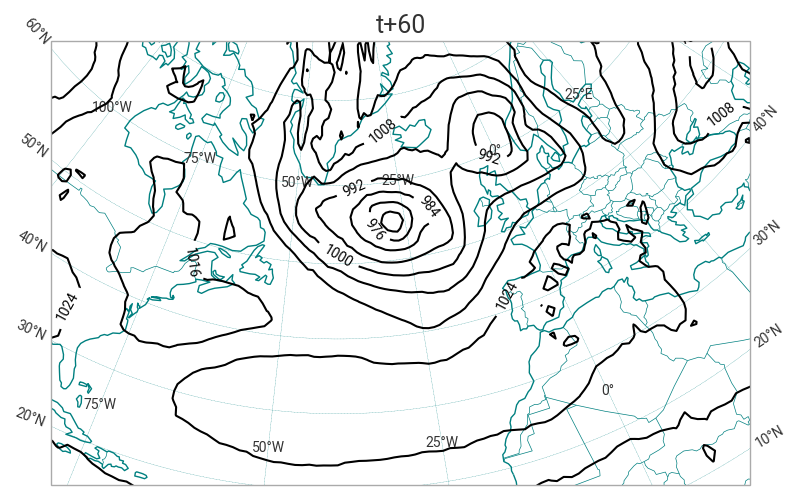}
    \end{minipage}
    \hfill
    \begin{minipage}{0.45\textwidth}
        \centering
        \includegraphics[width=\linewidth]{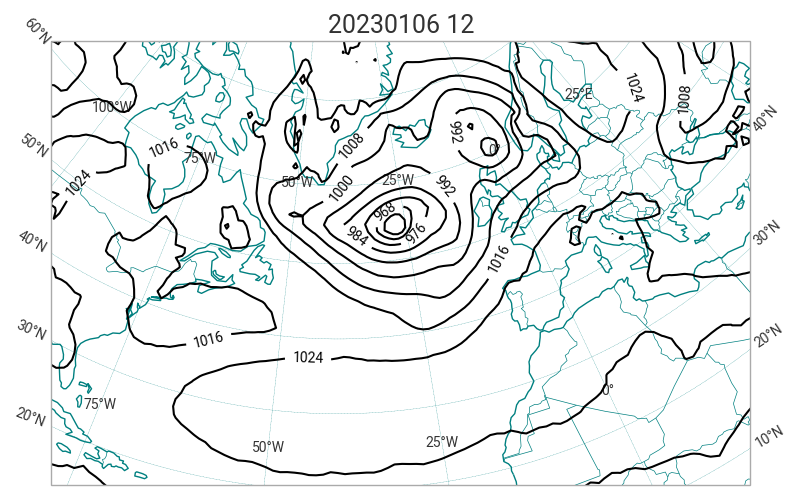}
    \end{minipage}

        \vspace{0.5cm}
    
    \begin{minipage}{0.45\textwidth}
        \centering
        \includegraphics[width=\linewidth]{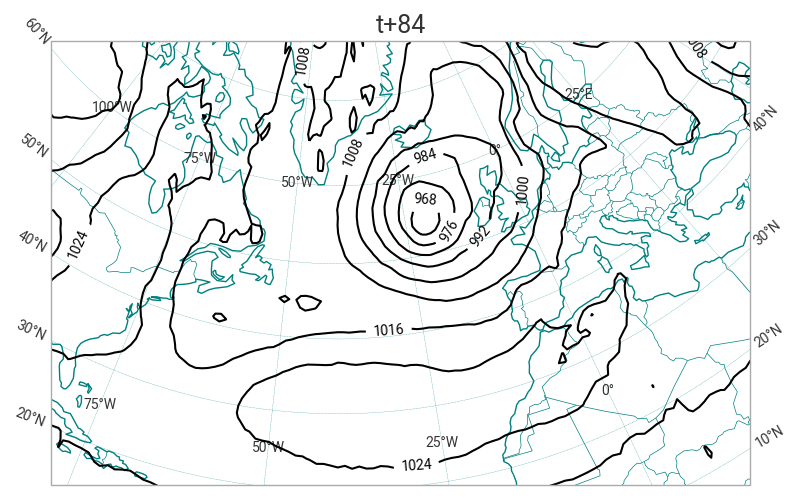}
    \end{minipage}
    \hfill
    \begin{minipage}{0.45\textwidth}
        \centering
        \includegraphics[width=\linewidth]{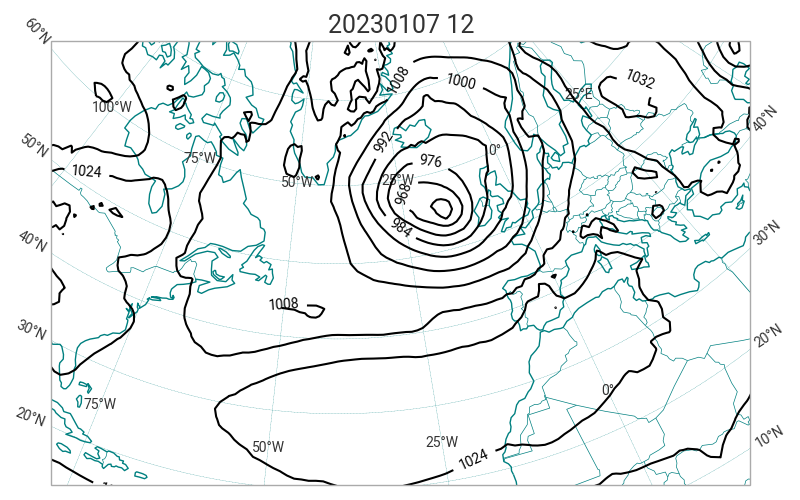}
    \end{minipage}

    \caption{Pressure at mean sea level [hPa] from GraphDOP predictions (left) and the operational ECMWF analysis (right) at 24-hour intervals, from a forecast initialised at 00UTC on January 4th, 2023.}
    \label{fig:learned_dynamics}
\end{figure}

\subsection{Evidence of learned atmospheric dynamics}
\label{subsec:learned_dynamics}

In this section we look at GraphDOP's predictions for a case study of explosive cyclogenesis, using the model trained with a more complete set of observations (as detailed in Table \ref{table:dop-observations}).

Figure \ref{fig:learned_dynamics} shows the evolution of mean sea level pressure in a single GraphDOP prediction initialised on January 4th, 2023. The right-hand panels show the operational ECMWF 4D-Var analysis valid at the time of each prediction. The case study features a rapidly intensifying extra-tropical cyclone developing from a small perturbation off the coast of Newfoundland on January 4th into a large depression with a central pressure below 960~hPa two days later; a drop of around 50~hPa. The GraphDOP prediction captures the genesis of this storm well in terms of both the location and intensity. A smaller, secondary low pressure system which formed just ahead of the main storm and passes close to Ireland on January 5th was also captured although the intensity was a little weaker in the GraphDOP prediction.

These results confirm that GraphDOP has learned representations of the complex dynamics of extra-tropical systems and in particular with respect to perturbation growth in a baroclinically unstable environment.

\section{Discussion and outlook}

The results in this paper, together with those in \cite{alexe2024}, suggest that machine learning models trained purely on observations of the Earth System develop internal representations of the physical world, including the Earth System state and dynamics as well as the characteristics of different observing systems. In many ways, the phrase "world model" used in other areas of machine learning is particularly suitable in this context.

Although the observation-driven approach marks a clear departure from the development arc that started with \cite{richardson1922} (and eventually led to today's analysis-driven ML models trained on the output of state-of-the-art NWP systems), many parallels can be drawn with traditional NWP.

Both traditional NWP and observation-driven neural networks have abstract internal representations of the physical world; the former are explicitly formulated based on physical laws and empirical heuristics, while the latter are emergent properties learned via gradient descent.

We have presented evidence that, like data assimilation, GraphDOP is able to combine indirect information from diverse irregularly spaced observations into a unified representation of the underlying Earth System state (Section \ref{sec:unified_state}). Learned "observation operators" naturally emerge, mapping from the model's internal latent state to the observed variables, capturing complex geometric and instrument-specific effects like sunglint and limb effects (Section \ref{sec:viewing_effects}). We have shown that GraphDOP is able to model the complex dynamics of extra-tropical weather systems with a skill that is starting to approach that of physics-based systems (Section \ref{subsec:learned_dynamics}).

The ability to infer structures that are not directly observed also has parallels with how a trained human meteorologist can look at a geostationary satellite image showing cloud patterns over the North Atlantic and, using a combination of experience and an internal conceptual model of the structure and dynamics of extra-tropical weather systems, can correctly infer the location of the lowest mean sea level pressure alongside estimates of the likely path and strength of the jet stream.

One area where the two approaches differ relates to their use of prior knowledge. Data assimilation combines information from new observations with prior knowledge in a Bayesian framework. This prior knowledge comes in the form of a background estimate of the current state and a physical model which defines how variables are related in both time and space. Although GraphDOP was given no prior knowledge at the beginning of the training, at inference time we can consider the relationships it has acquired as learned priors. For example, in the absence of observations we have seen evidence that the predictions revert to a learned climatological prior. Similarly, we have seen signs of structural and dynamical priors (a learned dynamical model) which it uses to infer unobserved and non-local variables (Section \ref{sec:struct_dynamics}).

One of the main limitations of the current GraphDOP configuration is the restriction of only using 12-hours of observations as input, with no background state provided. To address this we plan to introduce recurrence and cycle through several input windows carrying information forward via the latent space. This approach will add a new prior estimate of the current state as an additional input alongside the observations. The network should then learn to combine this prior with the incoming observations in a way that it finds to be optimal at minimising the forecast error.

Other avenues for future research include investigating the model's handling of systematic observation errors. It remains an open question whether external bias correction schemes are necessary, or if the model can learn and correct for these biases internally when provided with a sufficiently diverse set of observations. Questions also remain over whether the observation coverage is sufficiently complete to provide enough constraint for observation-only approaches to reach their potential.

In our opinion, the results from the GraphDOP model presented in \cite{alexe2024} and in this paper are highly encouraging. While results are still improving, it still remains to be seen how far these approaches can be taken and the extent to which they will ultimately challenge the state-of-the-art. We hope that this work establishes a firm foundation for further research in this exciting direction.

\section{Acknowledgements}
Many people have contributed in various ways to AI-DOP at ECMWF. We would like to thank Christian Lessig, Mat Chantry, Peter Dueben, Chris Burrows, Sean Healy, Patricia de Rosnay, Mike Rennie, Florian Pinault, Mario Santa Cruz, Cathal O'Brien, Jan Polster, Aaron Hopkinson, Marcin Chrust and Alan Geer. We acknowledge the Partnership for Advanced Computing in Europe (PRACE) for awarding us access to Leonardo, CINECA, Italy.

\bibliographystyle{plainnat}
\bibliography{references} 

\newpage
\appendix

\section*{Appendix}

\section{Specification of training datasets}

As used in Sections \ref{sec:unified_state}, \ref{sec:viewing_effects} and \ref{subsec:learned_dynamics}.

\begin{table}[H]
\centering
\begin{tabular}{|l|l|c|l|}
\hline
\textbf{Category} & \textbf{Instrument} & \textbf{Period} & \textbf{Parameters} \\
\hline
\multirow{8}{*}{Microwave Sounders} 
& NPP ATMS & 2013-2023 & channels 1-22 \\
& NOAA 20 ATMS & 2018-2023 & channels 1-22 \\
& NOAA 15 AMSU-A & 2013-2023 & channels 1-15 \\
& NOAA 18 AMSU-A & 2013-2023 & channels 1-15 \\
& METOP-A AMSU-A & 2013-2021 & channels 1-15 \\
& METOP-B AMSU-A & 2013-2023 & channels 1-15 \\
& NOAA 19 MHS    & 2009-2023 & channels 1-5 \\
& METOP-B MHS    & 2013-2023 & channels 1-5 \\
\hline
\multirow{1}{*}{Infrared Sounders}
& METOP-B IASI & 2013-2023 & 17 channels \\
\hline
\multirow{1}{*}{Visible}
& METOP-B AVHRR & 2013-2023 & visible channel \\
\hline
\multirow{10}{*}{Conventional - surface}
& Automatic Land SYNOP & 2013-2023 & ps, t2m, rh2m, u10, v10 \\
& Manual Land SYNOP & 2013-2023 & ps, t2m, rh2m, u10, v10, cloud fraction \\
& BUFR Land SYNOP & 2014-2023 & ps, t2m, rh2m, u10, v10, cloud fraction \\
& SHIP & 2013-2023 & ps, t2m, rh2m, u10, v10, cloud fraction \\
& BUFR SHIP SYNOP & 2014-2023 & ps, t2m, rh2m, u10, v10, cloud fraction \\
& Abbreviated SHIP & 2013-2023 & ps, t2m, rh2m, u10, v10 \\
& METAR & 2013-2023 & ps, t2m, rh2m, u10, v10, cloud fraction \\
& Automatic METAR & 2013-2023 & ps, t2m, rh2m, u10, v10 \\
& DRIBU & 2013-2023 & ps, sst \\
& BUFR Drifting Buoys & 2016-2023 & ps, sst \\
\hline
\multirow{5}{*}{Conventional - sonde}
& TEMP SHIP & 2013-2023 & z, t, u, v on standard pressure levels \\
& BUFR SHIP TEMP & 2014-2023 & z, t, u, v on standard pressure levels \\
& Land TEMP & 2013-2023 & z, t, u, v on standard pressure levels \\
& BUFR Land TEMP & 2014-2023 & z, t, u, v on standard pressure levels \\ 
& Dropsondes & 2013-2023 & z, t, u, v on standard pressure levels \\
\hline
\multirow{1}{*}{Other}
& Atmospheric Motion Vectors & 2013-2023 & u,v \\
\hline
\end{tabular}
\caption{Input and output Earth System observations used in this study.}
\label{table:dop-observations}
\end{table}

\newpage
\section{SEVIRI-only experiment details}
\label{appendix:geos_exp_setup}

As used in Section \ref{subsec:structure}.

\begin{table}[H]
\centering
\begin{tabular}{|l|l|c|l|l|}
\hline
\textbf{Input} & \textbf{output} & \textbf{Period} & \textbf{Parameters} & \textbf{Input/Output}\\
\hline
\multirow{2}{*}{Infrared Geostationary}
& Meteosat 10 SEVIRI & 2013-2018 & 5 IR channels & Input + Output\\
& Meteosat 11 SEVIRI & 2018-2023 & 5 IR channels & Input + Output\\
\hline
\multirow{10}{*}{Conventional - surface}
& Automatic Land SYNOP & 2013-2023 & ps, t2m, rh2m, u10, v10 & Output\\
& Manual Land SYNOP & 2013-2023 & ps, t2m, rh2m, u10, v10, cf & Output\\
& BUFR Land SYNOP & 2014-2023 & ps, t2m, rh2m, u10, v10, cf & Output\\
& SHIP & 2013-2023 & ps, t2m, rh2m, u10, v10, cf & Output\\
& BUFR SHIP SYNOP & 2014-2023 & ps, t2m, rh2m, u10, v10, cf & Output\\
& Abbreviated SHIP & 2013-2023 & ps, t2m, rh2m, u10, v10 & Output\\
& METAR & 2013-2023 & ps, t2m, rh2m, u10, v10, cf & Output\\
& Automatic METAR & 2013-2023 & ps, t2m, rh2m, u10, v10 & Output\\
& DRIBU & 2013-2023 & ps, sst & Output\\
& BUFR Drifting Buoys & 2016-2023 & ps, sst & Output\\
\hline
\multirow{5}{*}{Conventional - sonde}
& TEMP SHIP & 2013-2023 & z, t, u, v on standard pressure levels & Output\\
& BUFR SHIP TEMP & 2014-2023 & z, t, u, v on standard pressure levels & Output\\
& Land TEMP & 2013-2023 & z, t, u, v on standard pressure levels & Output\\
& BUFR Land TEMP & 2014-2023 & z, t, u, v on standard pressure levels & Output\\ 
& Dropsondes & 2013-2023 & z, t, u, v on standard pressure levels & Output\\
\hline
\multirow{2}{*}{Other}
& Atmospheric Motion Vectors & 2013-2023 & u,v & Output\\
& (all available satellites) & & & \\
\hline

\end{tabular}
\caption{Input and output observations used in the experiment in Section \ref{subsec:structure}.}
\label{table:geo-exp-observations}
\end{table}

\section{Instrument acronyms}
\label{appendix:acronyms}

\begin{table}[ht]
\centering
\caption{Instrument name definitions.}
\label{tab:instrument_names}
\begin{tabular}{ll}
\toprule 
Acronym & Full name \\
\midrule
AMSU-A & Advanced Microwave Sounding Unit-A \\
ATMS   & Advanced Technology Microwave Sounder \\
AVHRR  & Advanced Very High Resolution Radiometer \\
IASI   & Infrared Atmospheric Sounding Interferometer \\
MHS    & Microwave Humidity Sounder \\ 
SEVIRI & Spinning Enhanced Visible and Infrared Imager \\
\bottomrule
\end{tabular}
\end{table}

\end{document}